\documentclass[twocolumn, secnumarabic, amssymb, nobibnotes, aps]{revtex4-1}
\usepackage{amsmath,amstext,bm,appendix,graphicx,multirow,float,slashed}
\usepackage{color}
\usepackage{diagbox}
\usepackage{subfigure}
\usepackage{color}
\usepackage[bookmarks=false]{hyperref}
\hypersetup{colorlinks=true,citecolor=blue,linkcolor=blue,urlcolor=blue,pdfstartview=FitH,bookmarksopen=true}
\definecolor{green}{rgb}{0.8,0.98,0.83}
\pagecolor{white}

\begin{document}

\title{Microwave Amplification in a $\mathcal{PT}$-symmetric-like Cavity Magnomechanical System}
\author{Hua Jin$^{1}$}%
\author{Zhi-Bo Yang$^{1}$}%
\author{Jing-Wen Jin$^{1}$}%
\author{Jian-Yu Liu$^{1}$}%
\author{Hong-Yu Liu$^{1}$}%
\email{liuhongyu@ybu.edu.cn}
\author{Rong-Can Yang$^{2,3,4}$}%
\email{rcyang@fjnu.edu.cn}
\affiliation{$^{1}$Department of Physics, College of Science, Yanbian University, Yanji, Jilin 133002, China}
\affiliation{$^{2}$College of Physics and Energy, Fujian Normal University, Fujian Provincial Key Laboratory of Quantum Manipulation and New Energy Materials, Fuzhou, 350117, China}
\affiliation{$^{3}$Fujian Provincial Engineering Technology Research Center of Solar Energy Conversion and Energy Storage, Fuzhou, 350117, China}
\affiliation{$^{4}$Fujian Provincial Collaborative Innovation Center for Advanced High-Field Superconducting Materials and Engineering, Fuzhou, 350117, China}

\date{\today}%
\begin{abstract}
We propose a scheme that can generate tunable magnomechanically induced amplification in a double-cavity parity-time-($\mathcal{PT}$-) symmetric-like magnomechanical system under a strong control and weak probe field. The system consists of a ferromagnetic-material yttrium iron garnet (YIG) sphere placed in a passive microwave cavity which is connected with another active cavity. We reveal that ideally induced amplification of the microwave probe signal may reach the maximum value $10^6$ when cavity-cavity, cavity-magnon and magnomechanical coupling strengths are nonzero simultaneously. 
The phenomenon might have potential applications in the field of quantum information processing and quantum optical devices. 
Besides, we also find the phenomena of slow-light propagation.
In this case, group speed delay of the light can achieve $3.5\times10^{-5}s$, which can enhance some nonlinear effect. 
Moreover, due to the relatively flat dispersion curve, the proposal may be applied to sensitive optical switches, which plays an important role in storing photons and quantum optical chips.
\end{abstract}
\maketitle

\section{introduction}
The interaction between light and matter is an important subject in the field of quantum optics. 
The study of light toward the perspective of quantum leads to some interesting phenomena different from classical ones. 
One of the most famous phenomenon is induced transparency (such as electromagnetically/optomechanically induced transparency)~\cite{001,002,003,004,005,006,007,008,009,010,011,012,013,014,015,016}, as well as induced absorption~\cite{003,014} and induced amplification~\cite{015,016,017,018,019,020} which has been widely studied in current decades. 
Besides, signal amplification whose aim is to  increase signal-to-noise ratio is significantly crucial in the field of quantum information and quantum optics. 
It is known that optical amplification usually results from the inversion of particle numbers under the action of a pumping field and stimulated radiation. 
It can directly amplify optical signals without converting them into electrical ones so as to possess a high degree of transparency on the format and rate of signals, making the whole optical fiber communication transmission system more simple and flexible~\cite{015,016}. 
It is noted that there are many mechanisms of light amplification, such as adding external drive and changing detuning conditions. Through the coupling effect of strong photon tunneling, double-cavity OMS not only shows the characteristics of photomechanically-induced absorption, photomechanically-induced amplification and simple normal mode splitting (NMS), but also adjusts the photon tunneling intensity.  The transformation from photomechanical induced absorption to photomechanical induced amplification can be further realized.  In this article we build our mechanism by adding active cavities.  In addition, the added gain scheme is widely used in quantum information and quantum communication due to its excellent characteristics of convenience and easy adjustment and may very useful for optical and microwave amplifiers~\cite{021}.

Parity-time($\mathcal{PT}$) symmetry, the non-Hermitian Hamitonian, which has a real spectra was proposed by Bebder in 1998 firstly and attracted wide attention ~\cite{022,023,024,025,026,027,028,029,030}. Since $\mathcal{PT}$-symmetry requires a strict balance between loss and gain. However the balance condition may be too difficult for the realistic implementation, especially when tiny disturbances are inevitable. $\mathcal{PT}$-symmetric-like system not requiring the strict balance can still follow the predictions of the $\mathcal{PT}$-symmetry in many cases and thus attract considerable attention~\cite{031,032}. At the exceptional point, where the system undergoes the transition from the $\mathcal{PT}$-symmetric-like phase and $\mathcal{PT}$-symmetric-like broken, pairs of eigenvalues collide and become complex has manifested in various physical system, such as photonics, electronics, acoustics, phononics. And the OMIA of the $\mathcal{PT}$-symmetric OMS has been achieved in the whispering-gallery-mode microtoroidal cavities. Common PT symmetric systems have two-cavity systems, but there are also examples of single cavities achieving effective gain by introducing external drives or other means  ~\cite{027}.

In the past few years, cavity magnonics, a new interdiscipline, attracted much attention. 
It mainly explores the interaction between confined electromagnetic fields and magnons, especially Yttrium iron garnet (YIG)~\cite{033,034,035,036,037,038,039,040,041}.
The reason is that the Kittel mode within YIG has a low damping rate and holds great magnonic nonlinearities~\cite{039}.  
In addition, the high spin density of magnons allows strong coupling between magnons and photons, giving rise to quasiparticles, i.e. the cavity-magnon polaritons. Then strong coupling between magnons and cavity photons can be observed at both low and room temperature. 
In this case, a large number of quantum-information-related problems have been studied by this method, including the coupling of magnons with superconducting qubits, observation of bistability~\cite{042,043}, cavity spintronics, energy level attraction of cavity magnetopolaron, magnon dark modes. 
Other interesting phenomena including magneton-induced transparency (MIT), magnetically induced transparency (MMIT), and magnetically controlled slow light have also been studied~\cite{044}.

In this paper, we utilize a cavity-magnomechanical system, which consists of a YIG sphere placed inside a three-dimensional microwave cavity that is connected with an passive cavity to realize microwave amplification. Through the discussion the properties of absorption and transmission, we obtain the amplification in the context of $\mathcal{PT}$-symmetric-like cavity magnomechanical system. 

The remaining parts are organized as follows. In Sec.~\ref{s2}, we introduce the model of our proposal. In Sec.~\ref{s2}, we plot the magnomechanically induced transparency window profiles. In Sec.~\ref{s3}, we explore magnomechanically induced amplification of the $\mathcal{PT}$-symmetric-like  cavity magnomechanical system and slow light propegation Sec.~\ref{s4}, we present the conclusion of our work.
\begin{figure}
	
	\centering
	\includegraphics[width=1\linewidth,height=0.18\textheight]{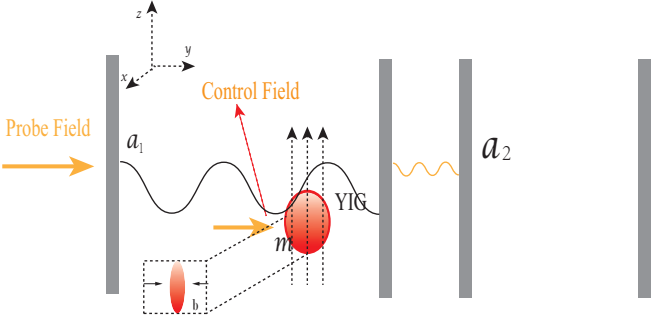}
	\hspace{0in}%
	\caption{Schematic of the setup studied in this paper. A cavity magnomechanical system consists of one ferromagnetic yttrium iron garnet (YIG) sphere placed inside a passive microwave cavity, which connected with an auxiliary cavity. A bias magnetic field is applied in the $z$ direction on the sphere to excites the magnon modes, which are strongly coupled with the cavity field. In the YIG sphere, bias magnetic field activates the magnetostrictive interaction. The magnetic coupling strength of a magnon depends on the diameter of the sphere and the direction of the external bias field~\cite{045}. We assumed that the YIG's magnomechanical interactions were directly enhanced by microwave driving (in the $y$ direction) its magnon mode. Cavity, phonon, and magnon modes are labeled $a_i, b, m$ ($i=1,2$).} 
	\label{fig1}
\end{figure}

\section{Model and Hamiltonian}\label{s2}
We use a hybrid cavity magnomechanical system that consists of one high-quality YIG sphere placed inside a microwave cavity which connects with another empty cavity, as shown in Fig.~\ref{fig1}. 
The YIG sphere has $250\mu{m}$ in diameter and ferric ions ${\rm Fe}^{+3}$ of density $\rho=4.22\times10^{27}m^{-3}$. 
This causes a total spin $S=5/2\rho{V}_m=7.07\times10^{14}$, where $V_m$ is the volume of the YIG and $S$ is the collective spin operator which satisfies the algebra i.e., $\left[S_{\alpha},S_{\beta} \right]=i\varepsilon^{\alpha\beta\gamma}S_{\gamma}$.
A uniform bias magnetic field (along z direction) is applied on the sphere, exciting the magnon mode that is then coupled to the first cavity field via magnetic-dipole interaction. 
In addition, the excitation of the magnon mode (i.e. Kittel mode) inside the sphere leads to a variable magnetization that results in the deformation of its lattice structure. 
The magnetostrictive force causes vibrations of the YIG, resulting in magnon-phonon interaction within YIG spheres~\cite{045}. 
It is noted that the single-magnon magnomechanical coupling strength depended on sphere diameter and direction of the external bias field is very weak. 
In this case, magnomechanical interaction of YIG can be enhanced by directly driving its magnon mode via an external microwave field. 
Furthermore, the first cavity is not only coupled to the second cavity, but also driven by a weak probe field. 
With consideration of the situation, the Hamiltonian for the whole system reads~\cite{044,046}
\begin{equation}\label{e001}
\begin{aligned}
\mathcal{H}/\hbar&=\omega_m\hat{m}^{\dag}\hat{m}+\omega_{a_1}\hat{a}^{\dag}_1\hat{a}_1+\omega_{a_2}\hat{a}^{\dag}_2\hat{a}_2+\omega_b\hat{b}^{\dag}\hat{b}\\
&+g_1(\hat{m}^{\dag}\hat{a}_1+\hat{m}\hat{a}^{\dag}_1)+g_2\hat{m}^{\dag}\hat{m}(\hat{b}+\hat{b}^{\dag})\\
&+J(\hat{a}^{\dag}_1\hat{a}_2+\hat{a}^{\dag}_2\hat{a}_1)+i\Omega(\hat{m}^{\dag}e^{-i\omega_{pu}t}-\hat{m}e^{i\omega_{pu}t})\\
&+i\varepsilon_{pr}(\hat{a}^{\dag}_1e^{-i\omega_{pr}t}-\hat{a}_1e^{i\omega_{pr}t}),\\
\end{aligned}
\end{equation}
where $ \hat{a}_j^{\dag} $($j=1,2$),  $ \hat{m}^{\dag} $and $ \hat{b}^{\dag} $ ($ \hat{a}_j $ , $\hat{ m} $ and $ \hat{ b} $) are the creation (annihilation) operators of the $ j $th cavity, magnon and phonon, respectively. They all satisfy the standard commutation relations for bosons. $ \omega_{a_j} $, $\omega_m $, $ \omega_b $ represent the resonance frequencies for the $ j $th cavity, magnon and phonon, respectively. $g_1 (J)$ denotes the coupling strength between the first cavity mode and magnon (the second cavity), and $g_2$ is the coupling constant between magnon and phonon. It is noted that the frequency $\omega_m$ is determined by the gyromagnetic ratio $\gamma$ and external bias magnetic field $H$ i.e., $\omega_m=\gamma H$ with $\gamma/2\pi=28\rm GHz$. In addition, $\Omega=\sqrt{5}/4\gamma \sqrt{N}B_0$ is the Rabi frequency, which is dependent of the coupling strength of the driving field with amplitude $B_0$ and frequency $\omega_{pu}$. And $\omega_{pr}$ is the probe field frequency having amplitude $\varepsilon_{pr}=\sqrt{2P_p\kappa_{1}/\hbar\omega_{pr}}$. It should be noted that we have ignored the nonlinear term $K\hat{m}^{\dag}\hat{m}^{\dag}\hat{m}\hat{m}$ in Eq.(1) that may arise due to strongly driven magnon mode~\cite{043} so as to ${K\left| \left\langle m\right\rangle \right| ^3}{\ll }\Omega$. With the rotating wave approximation, we can rewrite the whole Hamiltonian as 
\begin{equation}\label{e002}
\begin{aligned}
\mathcal{H}/\hbar&=\Delta_{m}\hat{m}^{\dag}\hat{m}+\Delta_{a_1}\hat{a}^{\dag}_1\hat{a}_1+\Delta_{a_2}\hat{a}^{\dag}_2\hat{a}_2+\omega_b\hat{b}^{\dag}\hat{b}\\
&+g_1(\hat{m}^{\dag}\hat{a}_1+\hat{m}\hat{a}^{\dag}_1)+g_2\hat{m}^{\dag}\hat{m}(\hat{b}+\hat{b}^{\dag})\\
&+J(\hat{a}^{\dag}_1\hat{a}_2+\hat{a}^{\dag}_2\hat{a}_1)+i\Omega(\hat{m}^{\dag}-\hat{m})\\
&+i\varepsilon_{pr}(\hat{a}^{\dag}_1e^{-i\delta t}-\hat{a}_1e^{i\delta t}),\\
\end{aligned}
\end{equation}
with $\Delta_{a_j}=\omega_{a_j}-\omega_{pu}$($j=1,2$), $\Delta_m=\omega_m-\omega_{pu}$, and $\delta=\omega_{pr}-\omega_{pu}$. 

In order to obtain the evolution of $a_j(t), m(t)$ and $b(t)$, we use quantum Heisenberg-Langevin equations, which can be expressed by
\begin{equation}\label{e003}
\begin{aligned}
\dot{\hat{a_1}}&=-i\Delta_{a_1}\hat{a}_1-ig_1\hat{ m}-\kappa_{1}\hat{a}_1+\varepsilon_{pr}e^{-i\delta t}\\
&+\sqrt{2\kappa_1}{\hat{a}_1}^{in}(t)-iJ\hat{a}_2,\\
\dot{\hat{a_2}}&=-i\Delta_{a_2}\hat{a}_2-\kappa_{2}\hat{a}_2+\sqrt{2\kappa_2}{\hat{a}_2}^{in}(t)-iJ\hat{a}_1,\\
\dot{\hat{m}}&=-i\Delta_m\hat{m}-ig_1\hat{a}_1-\kappa_m\hat{m}-ig_{2}\hat{m}(\hat{ b}+\hat{b}^{\dag})\\
&+\sqrt{2\kappa_m}\hat{m}^{in}(t)+\Omega,\\
\dot{\hat{b}}&=-i\omega_b\hat{b}-ig_2\hat{m}^{\dag}\hat{m}-\kappa_{b}\hat{ b}+\sqrt{2\kappa_b}\hat{b}^{in}(t)
\end{aligned}
\end{equation}
where $\kappa_1(\kappa_2),\kappa_b$ and $\kappa_m$ are the decay rates of the cavities, phonon and magnon modes, respectively. ${{\hat{a}_1}^{in}(t)}$, ${{\hat{a}_2}^{in}(t)}$,  ${\hat{b}^{in}(t)}$ and  ${\hat{m}^{in}(t)}$ are the vacuum input noise operators which have zero mean values and satisfies ${\left\langle \hat{q}^{in}\right\rangle}=0 (q=a_1, a_2, m, b) $. The magnon mode $m$ is strongly driven by a microwave field that causes a large steady-state amplitude corresponds to $ \arrowvert\langle m_s \rangle\arrowvert \gg 1 $. Moreover, owing to the magnon coupled to the cavity mode through the beam-splitter-type interaction, the two cavity fields also exhibit large amplitudes $ \arrowvert\langle a_{js} \rangle\arrowvert\gg 1 $. Then we can linearize the quantum Langevin equations around the steady-state values and take only the first-order terms in the fluctuating operator:${\left\langle\hat{O}\right\rangle}=O_s+{\hat{O}_+}{e^{-i\delta t}}+ {\hat{O}_-}{e^{i\delta t}}$~\cite{043}, where $\hat{O}=a_1, a_2, b, m.$ the steady-state solutions are given by
	\begin{equation}\label{e004}
	\begin{aligned}
	a_{1s}&=\frac{-(ig_1 m_s +iJa_{2s})}{i\Delta_{a_1}+\kappa_1}, a_{2s}=\frac{-iJa_{1s}}{i\Delta_{a_2}+\kappa_2},\\
	b_s&=\frac{-ig_{2}\left| m_s\right|^2 }{i\omega_b+\kappa_b},\\
	m_s&=\frac{-ig_1a_{1s}+\Omega}{i\widetilde{\Delta}_m+\kappa_m},\\
	\widetilde{\Delta}_m&=\Delta_m+g_2(b_s+{b_s}^*)\\
	\end{aligned}
	\end{equation}
 In order to achieve our motivation of signal amplification, we neglect off resonance terms to let $\hat{O}_-=0$, but $\hat{O}_+$ safisfying the relations
	
	\begin{equation}\label{e006}
	\begin{aligned}
	(i\lambda-\kappa_{1})\hat{{a}_1}_+-ig_1\hat{m}_+-iJ\hat{a_2}_++\varepsilon_{pr}&=0,\\
	(i\lambda-\kappa_{2})\hat{{a}_2}_+-iJ\hat{a_1}_+&=0,\\
	(i\lambda-\kappa_{m})\hat{m}_+-ig_1\hat{{a}_1}_+-iG\hat{b}_+&=0,\\
	(i\lambda-\kappa_{b})\hat{b}_+-iG^\ast\hat{m}_+&=0,\\
	\end{aligned}
	\end{equation}
where we have set $G=g_2m_s$, $\lambda=\delta-\omega_b$,  ${\omega_a}_i\gg\kappa_i$ $(i=1,2)$, and $\Delta_{a_1}=\Delta_{a_2}=\widetilde{\Delta}_m=\omega_b$. In this case, we can easily obtain
	
	\begin{equation}\label{e007}
	\begin{aligned}
	\hat{{a}_1}_+&=\frac{\varepsilon_{pr}}{\kappa_{1}-i\lambda+\frac{J^2}{\kappa_{2}-i\lambda}+\frac{{g_1}^2}{\kappa_{m}-i\lambda+\frac{\left|G \right|^2} {\kappa_{b}-i\lambda}}}.\\
	\end{aligned}
	\end{equation}
By use of the input-output relation for the cavity field $\varepsilon_{out}=\varepsilon_{in}-2\kappa_{1}\left\langle a_{1+} \right\rangle $ and setting $\varepsilon_{in}=0$, the amplitude of the output field can be written as
	\begin{equation}\label{e008}
	\begin{aligned}
	\varepsilon'_{out}&=\dfrac{\varepsilon_{out}}{\varepsilon_{pr}}=\frac{2\kappa_{1}\hat{{a}_1}_+}{\varepsilon_{pr}}.
	\end{aligned}
	\end{equation}
The real and imaginary parts of the output field are $\rm Re$ $[\varepsilon'_{out}]=\kappa_{1}(\hat{a}_{1+}+\hat{a}_{1+}^{*})/\varepsilon_{pr}$ and $\rm Im$ $[\varepsilon'_{out}]=\kappa_{1}(\hat{{a}}_{1+}-\hat{{{a}}}^*_{1+})/\varepsilon_{pr}$. 
These factors describe the absorption and dispersion of the systems, respectively.

\begin{figure}
	\centering
	\includegraphics[width=0.48\linewidth,height=0.16\textheight]{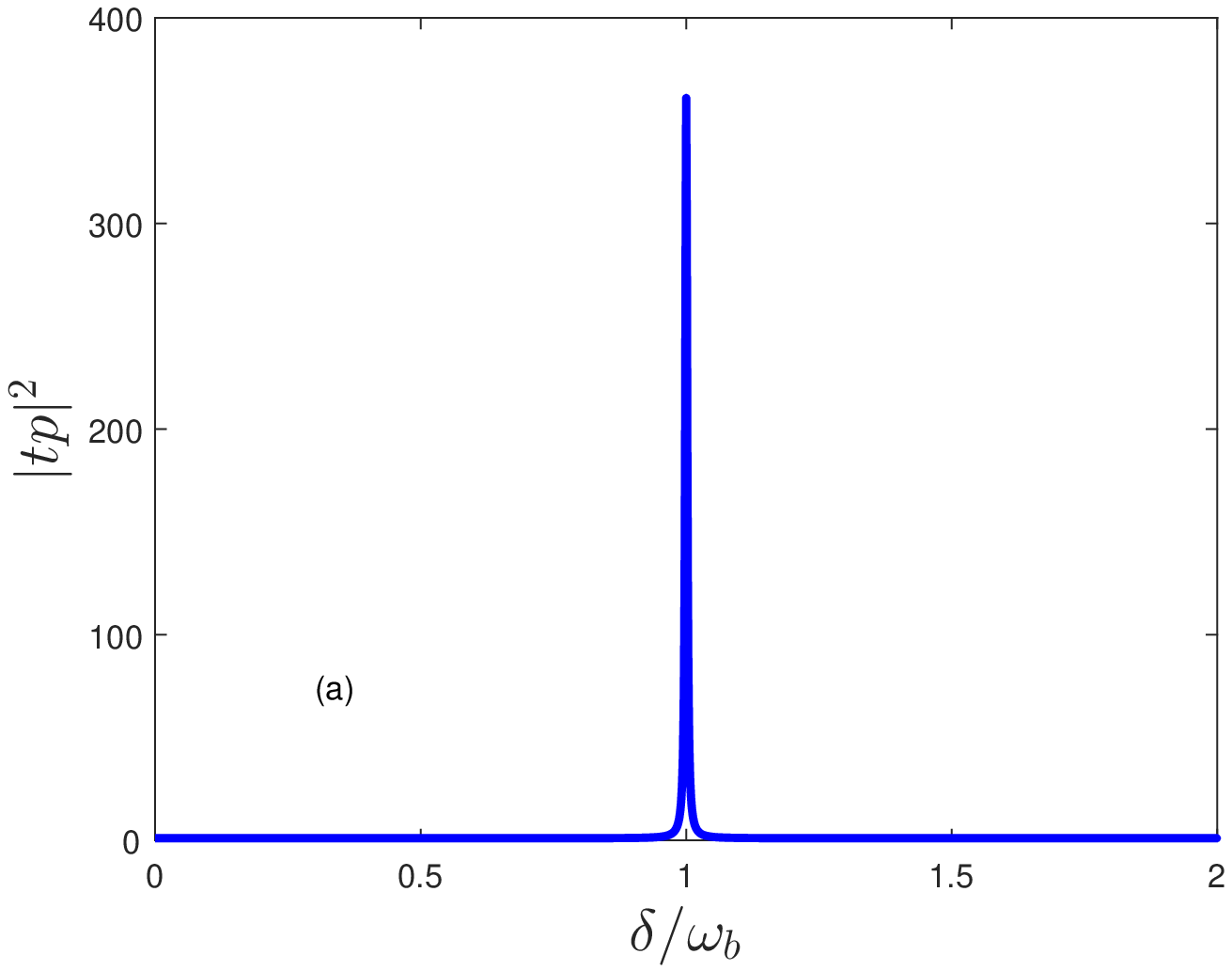}
	\hspace{0in}%
	\includegraphics[width=0.48\linewidth,height=0.16\textheight]{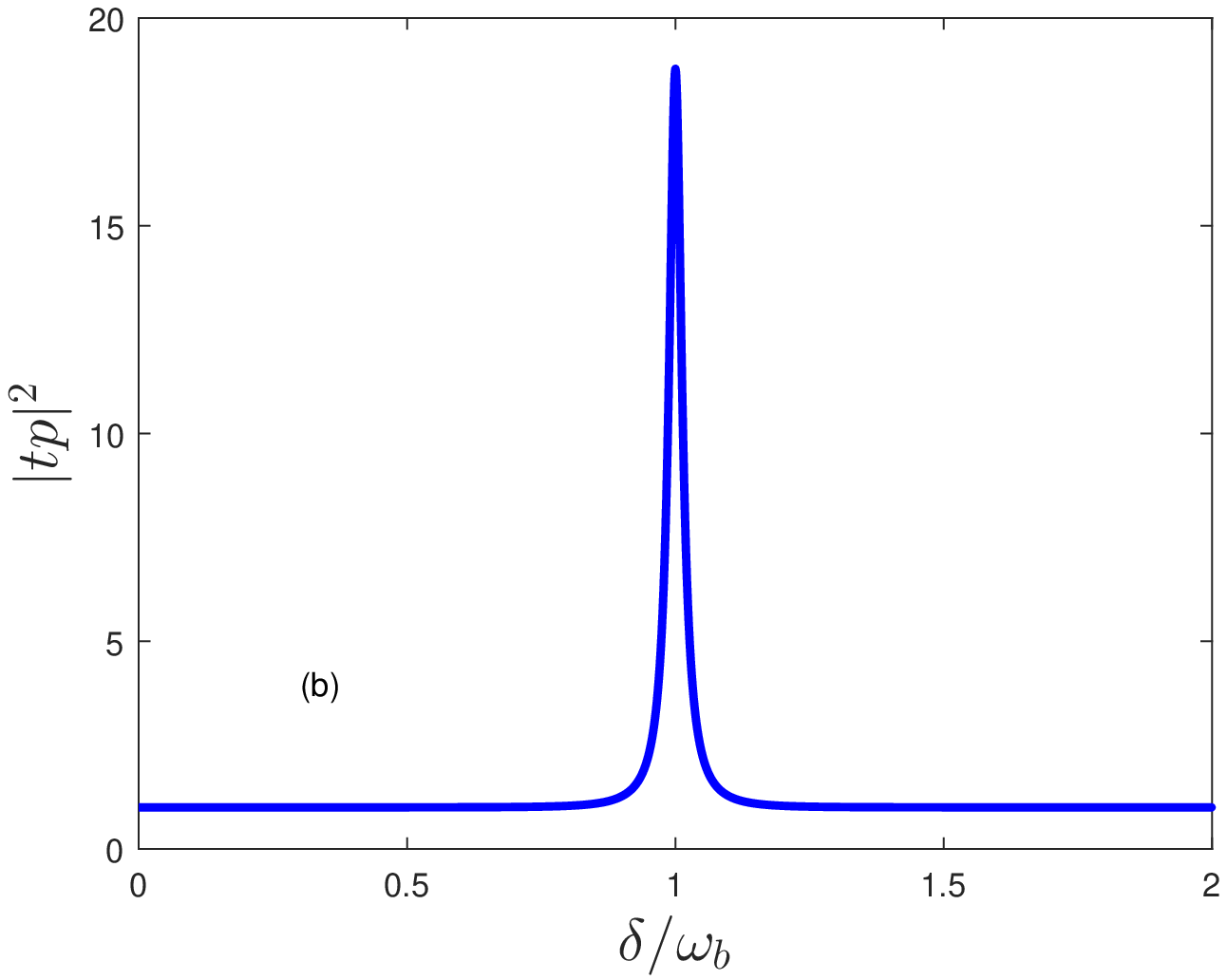}
	\hspace{0in}%
	
	\includegraphics[width=0.48\linewidth,height=0.16\textheight]{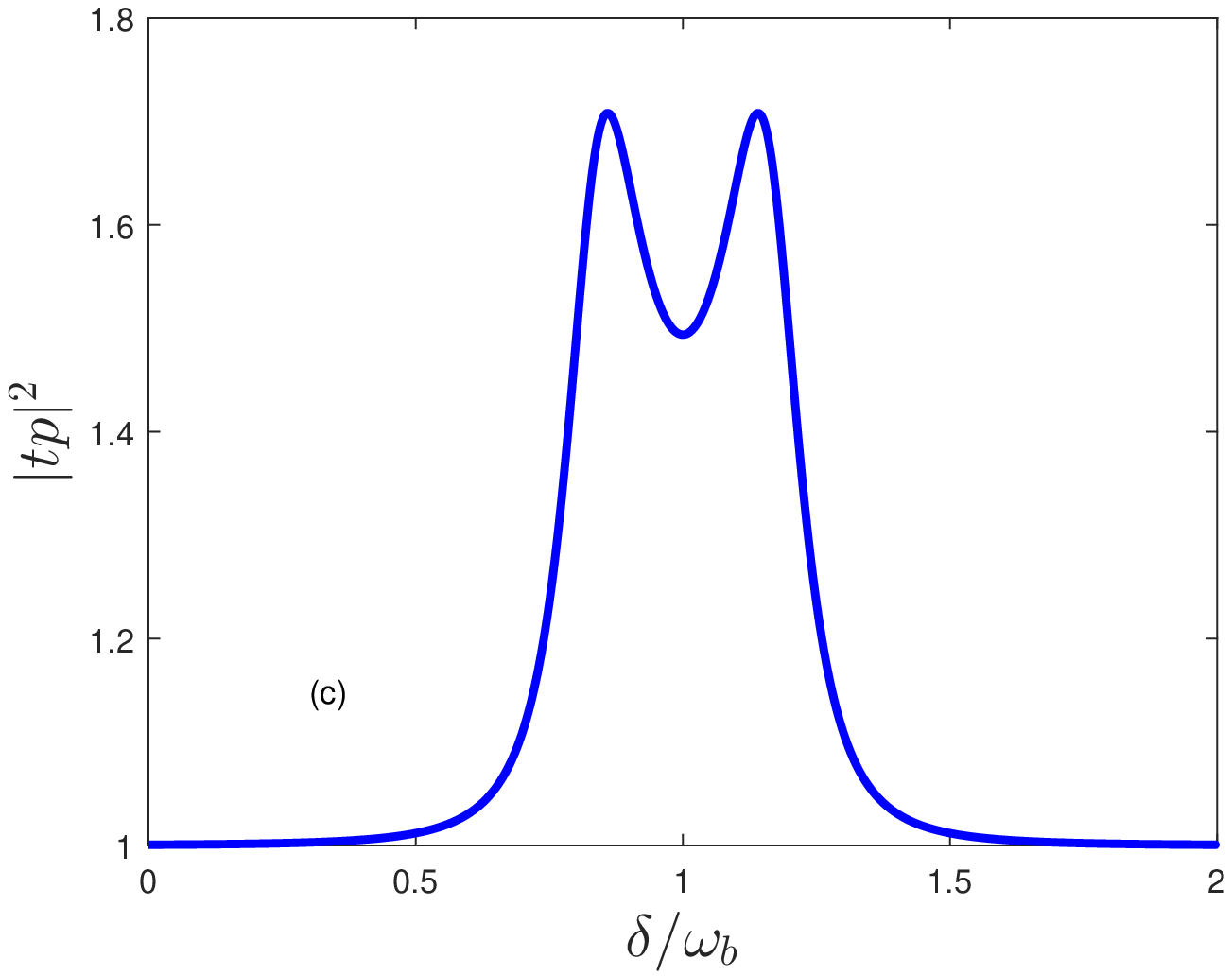}
	\hspace{0in}%
	\includegraphics[width=0.48\linewidth,height=0.16\textheight]{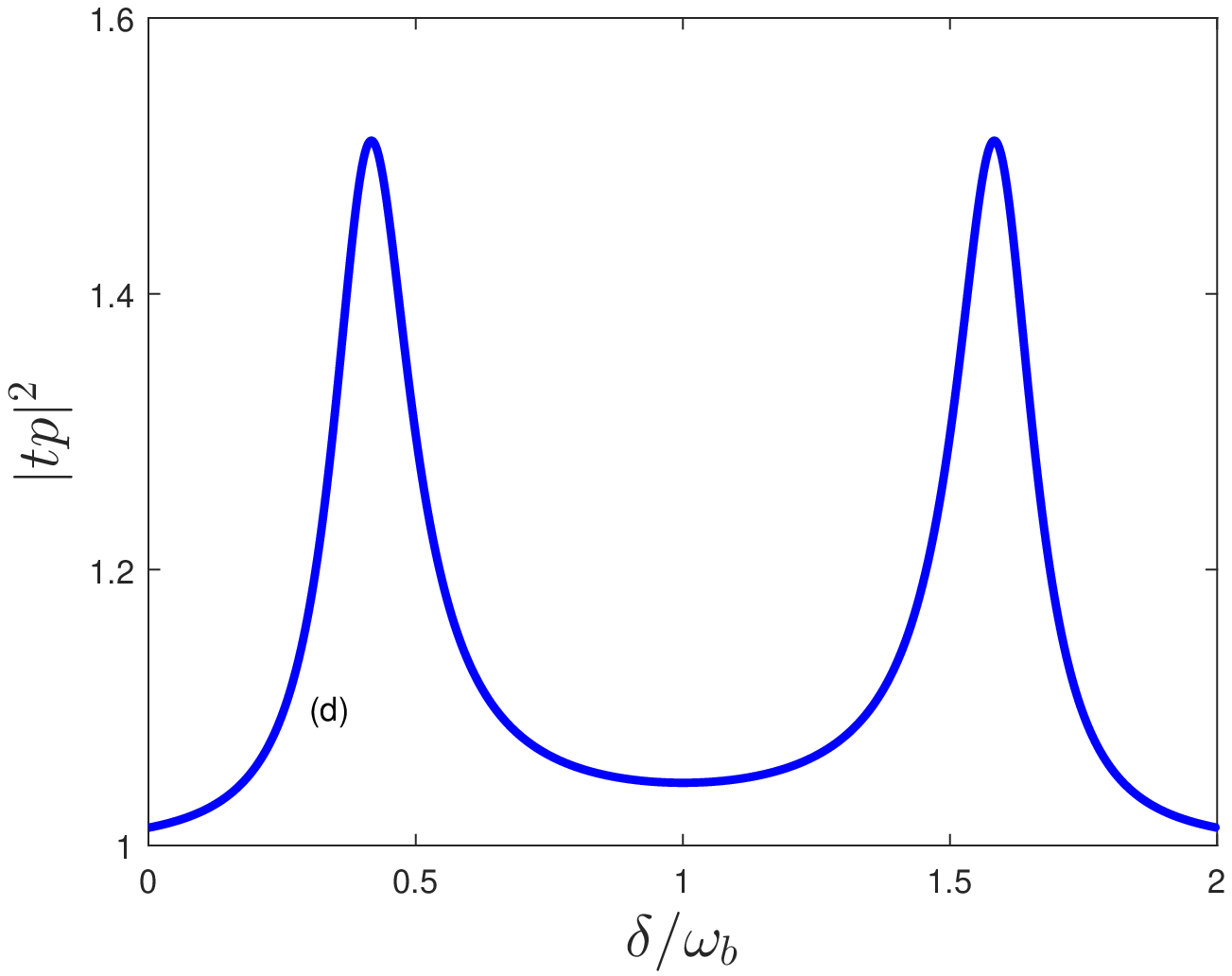}
	\hspace{0in}%
	\caption{The transmission $\left| t_p\right| ^2$  spectrum of probe field as function of $\delta/\omega_b$ when only interaction between two cavities is nonzero, (a)   $J/2\pi=0.6\rm MHz$, (b) $J/2\pi=0.8\rm MHz$, (c )$J/2\pi=2.0\rm MHz$ and (d) $J/2\pi=6\rm MHz$. }
	\label{fig2}
\end{figure}

\section{ Induced amplification and slow light propegation in $\mathcal{PT}$-symmetric-like magnomechanically systems} \label{s3}
For the numerical calculation, we use parameters chosen from a recent experiment on a  
hybrid magnomechanical system, where $\omega_{a_1}/2\pi=\omega_{a_2}/2\pi=10\rm {GHz}$, $\omega_b/2\pi=10\rm {MHz}$, $\kappa_b/2\pi=100\rm Hz$, $\omega_m/2\pi=10\rm GHz$, $\kappa_{1}/2\pi=2.0\rm MHz $, $\kappa_m/2\pi=0.1\rm MHz$, $g_1/2\pi=1.0\rm MHz$, $G/2\pi=3.5\rm MHz$,$\Delta_{a_1}=\Delta_{a_2}=\widetilde{\Delta}_m=\omega_b$, $\omega_d/2\pi=10\rm GHz$ are set ~\cite{026,033,034}.

At first, we consider the transmission rate $\left| t_p\right| ^2$ as a function of the probe detuning $\delta/\omega_b$ in the context of parity-time-($\mathcal{PT}$-) symmetric-like magnomechanical system. From Eq.~(\ref{e008}), the rescaled transmission  corresponding to the probe field can be expressed as

\begin{equation}\label{e009}
\begin{aligned}
t_p&=1-{\frac{2\kappa_{1}\hat{{a}_1}_+}{\varepsilon_{pr}}}.\\
\end{aligned}
\end{equation}

We first depict the transmission spectrum of the probe field against the scaled detuning $\delta/\omega_b$, for different values of $J$ in Fig.~\ref{fig2}, where the phonon-magnon coupling rate and photon-magnon interaction parameters are set to zero, i.e. ${G}={g_1}=0$. 
From  Fig.~\ref{fig2}(a), we can observe that the transmission peak near $\delta=\omega_b$ which is associated with the coupling rate of two cavities can much be larger than 1. 
The reason is that the gain cavity can scatter photons into the dissipative cavity. 
From Fig~\ref{fig2}(a)-(d), transmission coefficient decreases with the increase of coupling strength between two cavities. And we got a downward dip with two peaks From Fig~\ref{fig2}(c)-(d), amplification area becomes wider when $J$ getting lager simultaneously. This means that we can adjust the transmission coefficient and the size of the amplification region by changing the coupling between the two cavities when the system is double-cavity $\mathcal{PT}$-symmetric-like and the cavity contains no magnon.  

\begin{figure}
	\centering
	\includegraphics[width=0.48\linewidth,height=0.16\textheight]{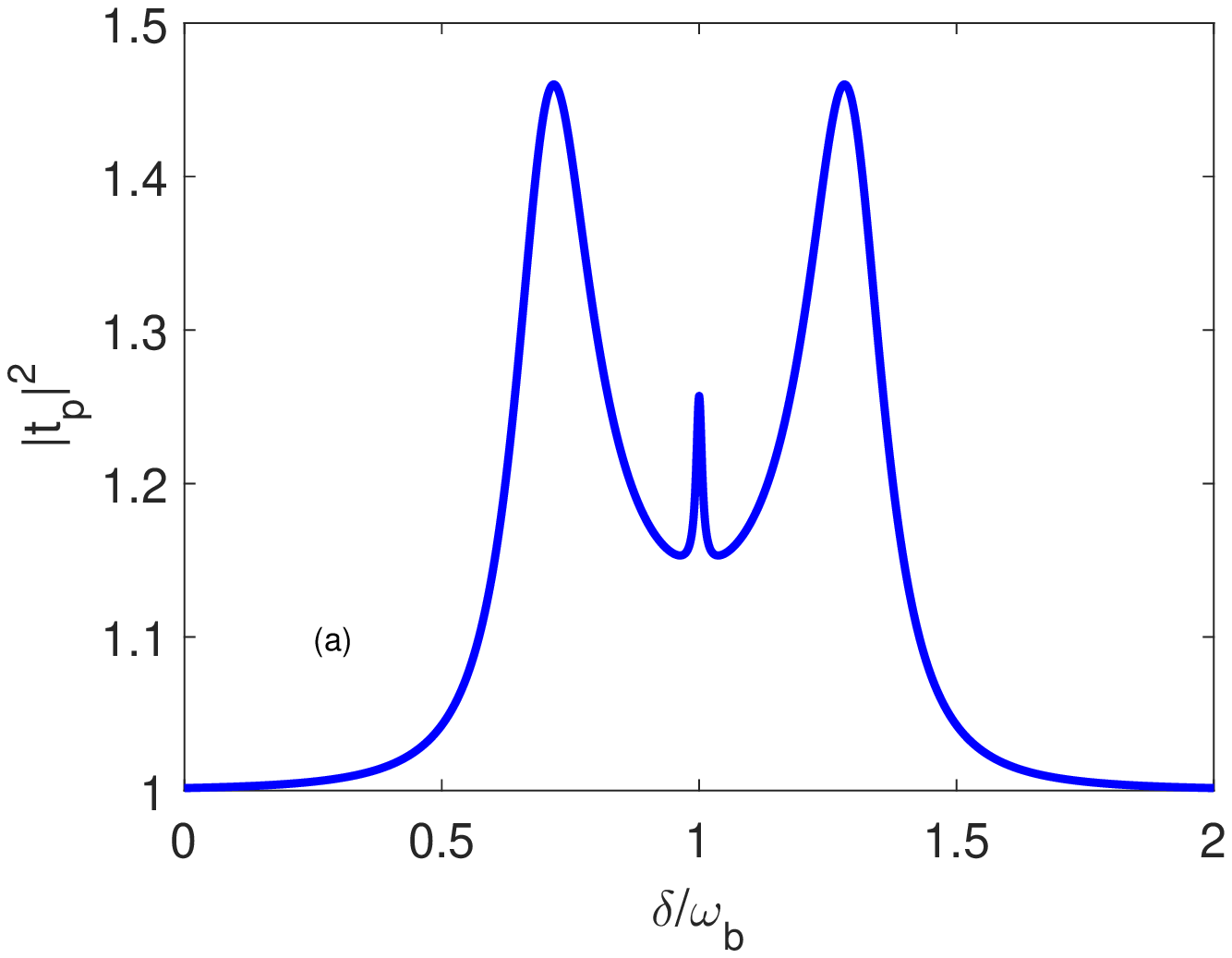}
	\hspace{0in}%
	\includegraphics[width=0.48\linewidth,height=0.16\textheight]{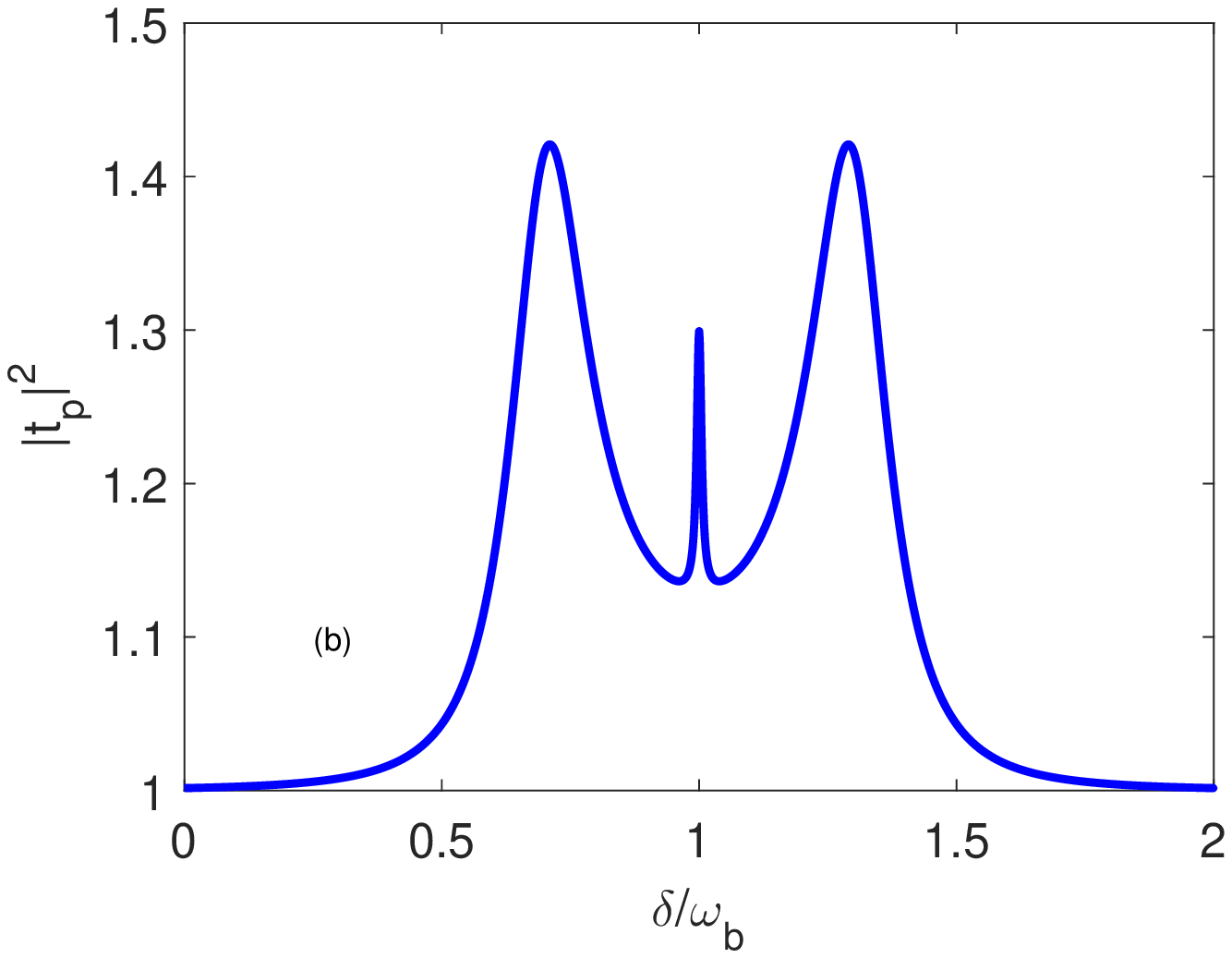}
	\hspace{0in}%

	\includegraphics[width=0.48\linewidth,height=0.16\textheight]{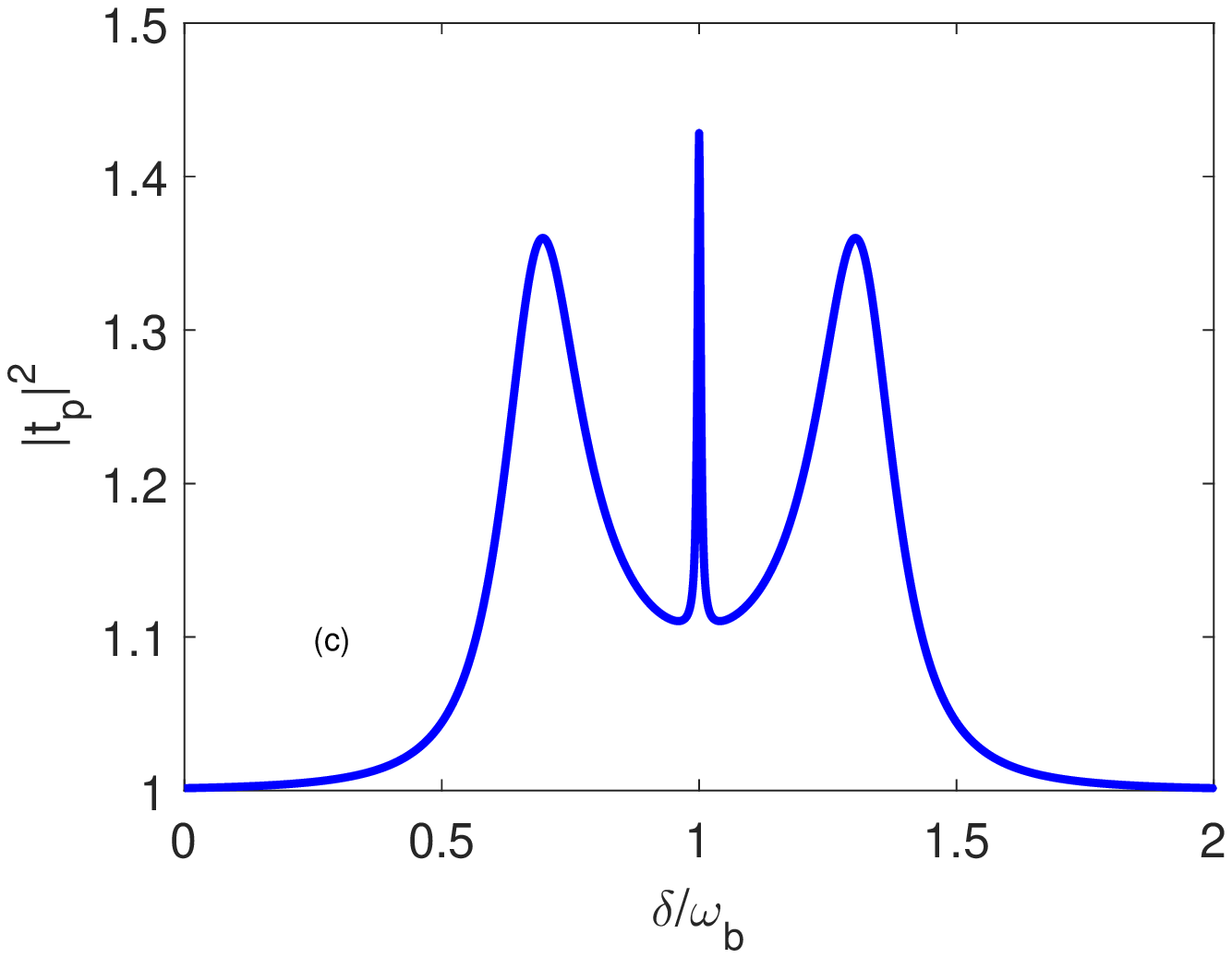}
	\hspace{0in}%
	\includegraphics[width=0.48\linewidth,height=0.16\textheight]{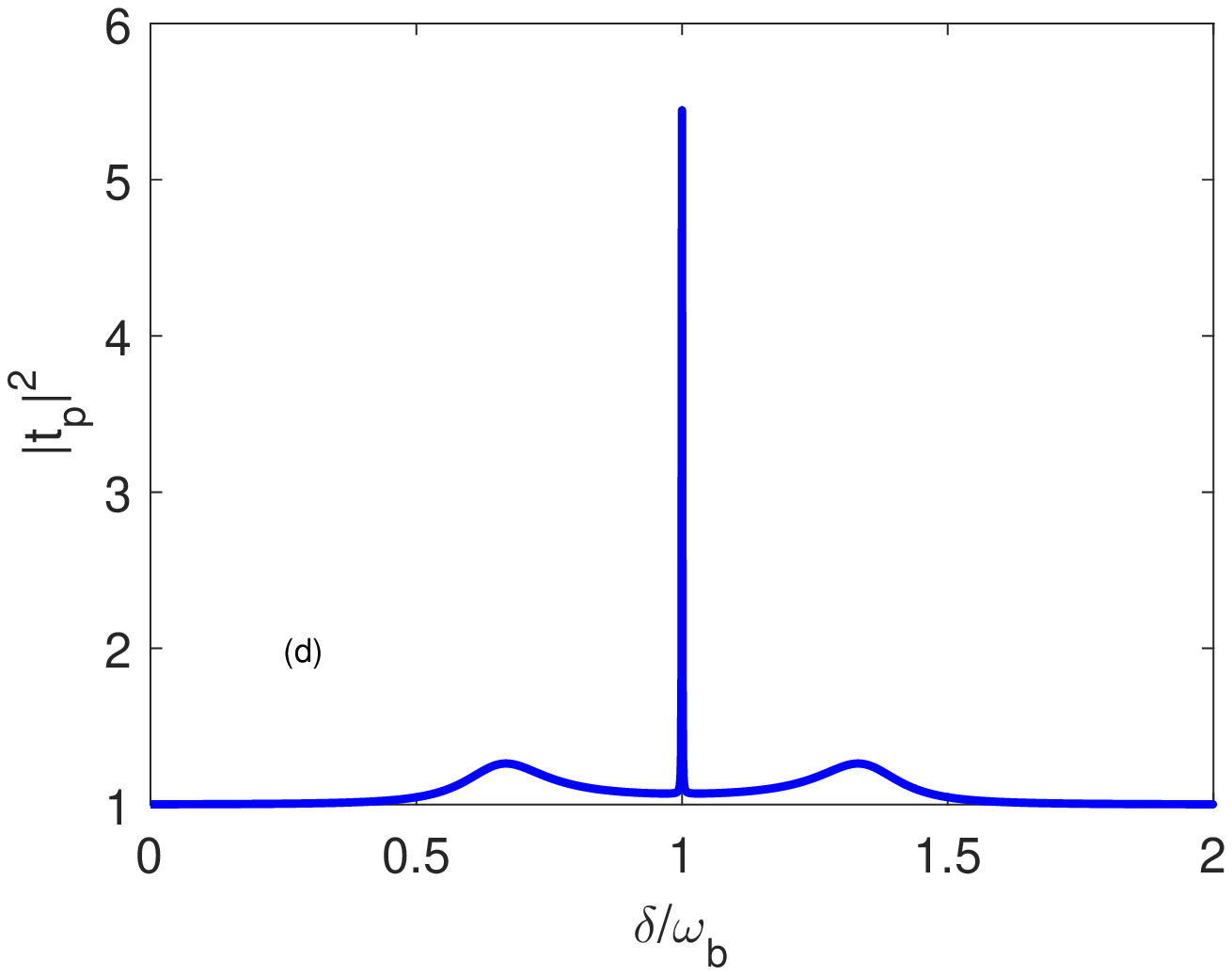}
	\hspace{0in}%
	\caption{The transmission $\left| t_p\right| ^2$  spectrum of probe field as function of $\delta/\omega_b$ when only coupling between magnon and phonon is absent means $G=0$, $J/2\pi=3.0\rm MHz$ (a) $g_1/2\pi=1.0\rm MHz$, (b) $g_1/2\pi=1.2\rm MHz$, (c) $g_1/2\pi=1.5\rm MHz$ and (d) $g_1/2\pi=2.0\rm MHz$. }
	\label{fig3}
\end{figure}

Next, we introduced one more coupling constant only set the coupling between magnon-phonon $G=0$. We got another peak near $\delta=\omega_b$ compared with Fig~\ref{fig2}(c)-(d), which was caused by coupling between magnon-photon in Fig~\ref{fig3}. This is because the magnon can scatter the photons of the active field into the probe field via indirect interaction. However, from Fig~\ref{fig3}(b)-(d), the middle peak became taller when $g_1$ increases. Hence the effect of light amplification caused by the interaction of magnon-photon get better as $g_1$ increase. And with the increasing of middle peak, the height of two peaks on both sides stay the same, that is, the light amplification caused by the coupling between two cavities not affected by $g_1$. However, the amplification effect is not ideal.

\begin{figure}
	\centering
	\includegraphics[width=0.48\linewidth,height=0.16\textheight]{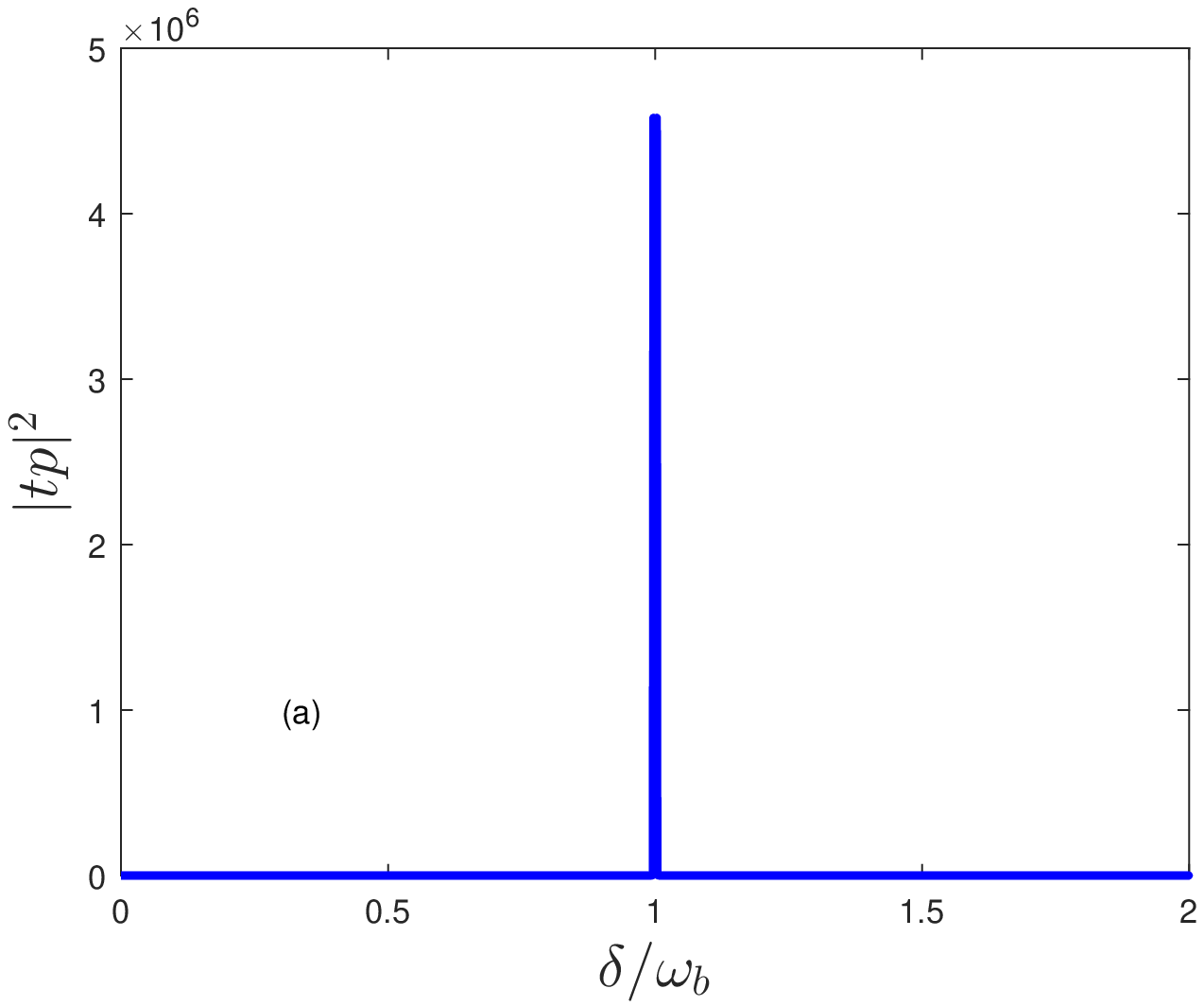}
	\hspace{0in}%
	\includegraphics[width=0.48\linewidth,height=0.16\textheight]{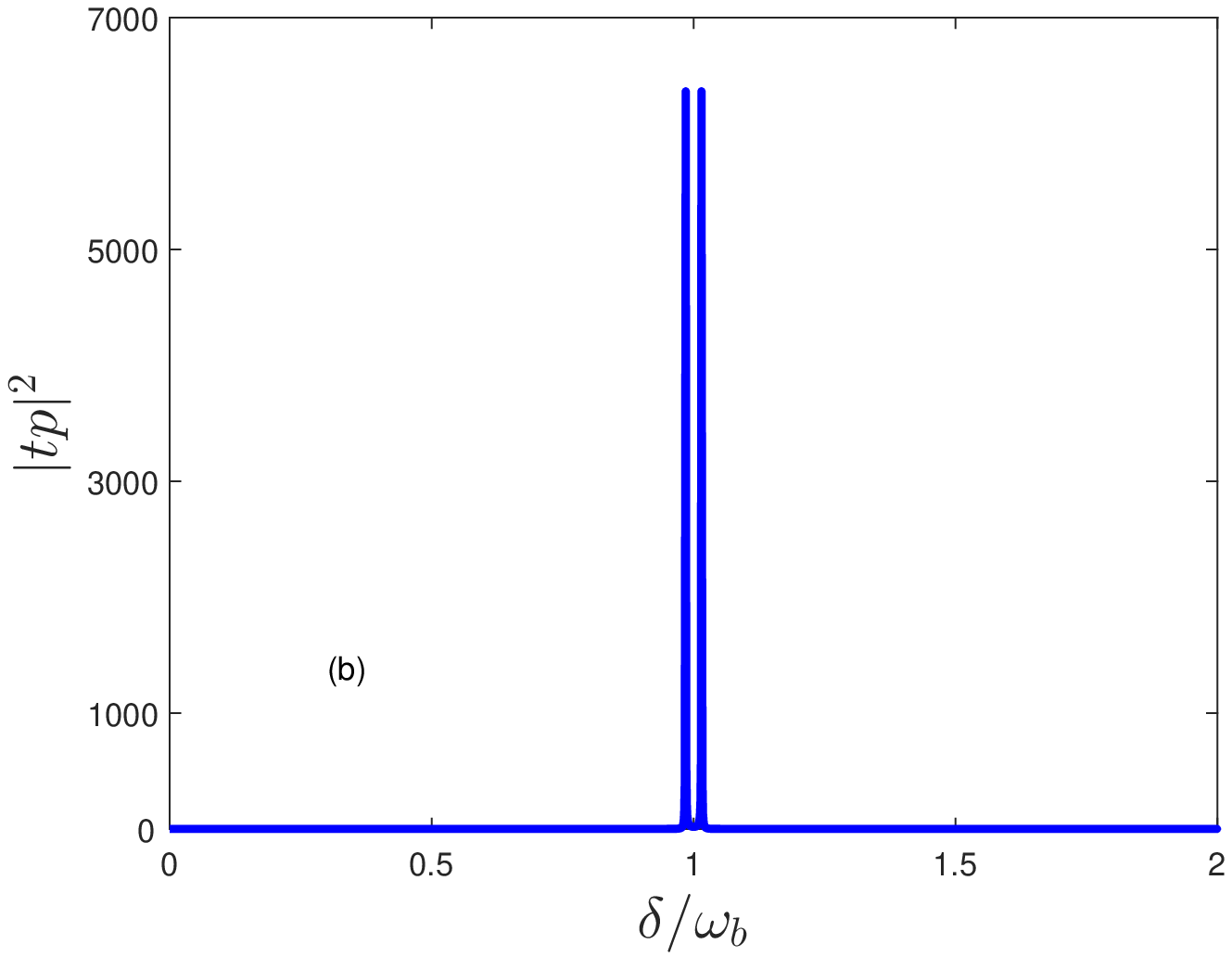}
	\hspace{0in}%
		\includegraphics[width=0.48\linewidth,height=0.16\textheight]{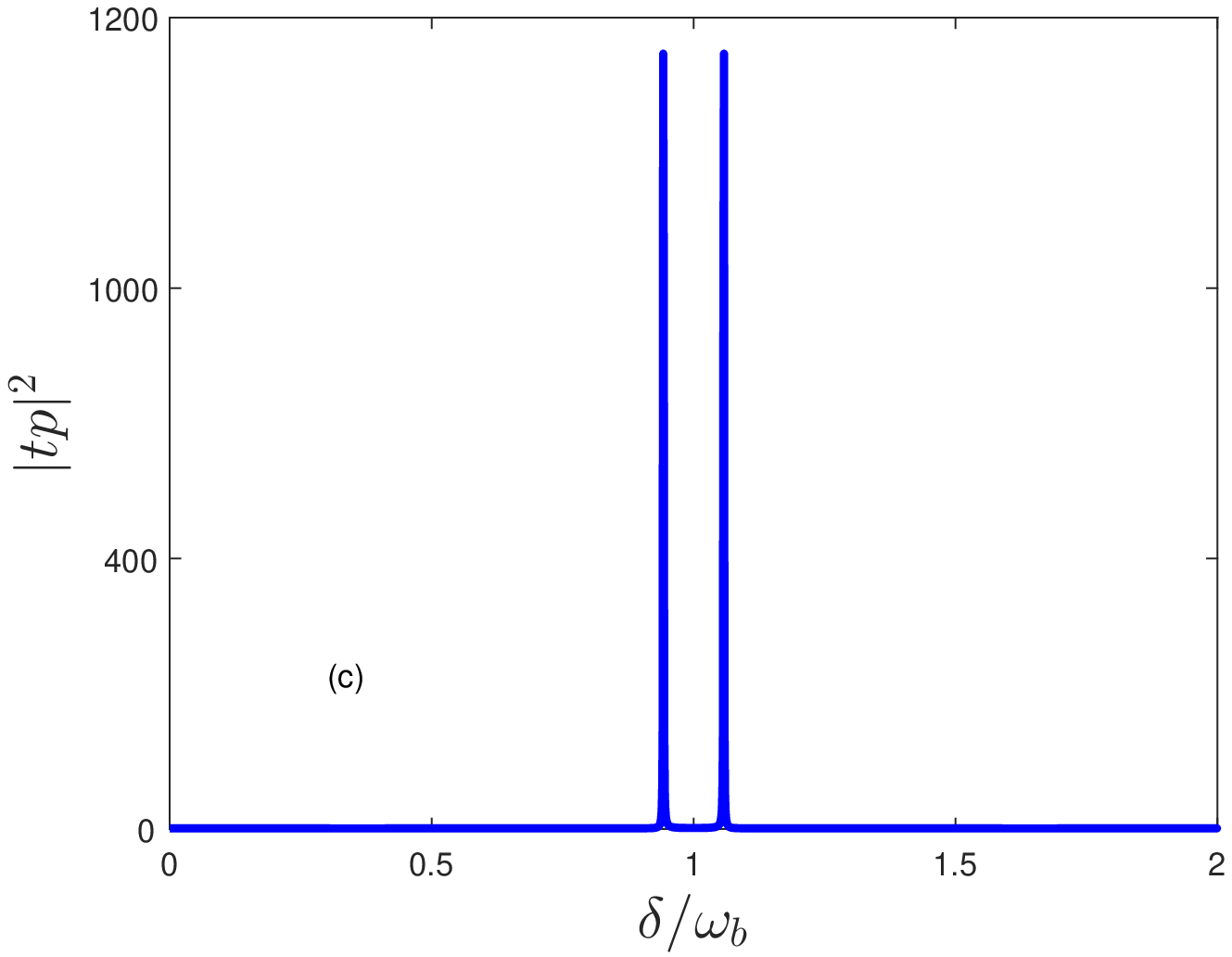}
	\hspace{0in}%
	\includegraphics[width=0.48\linewidth,height=0.16\textheight]{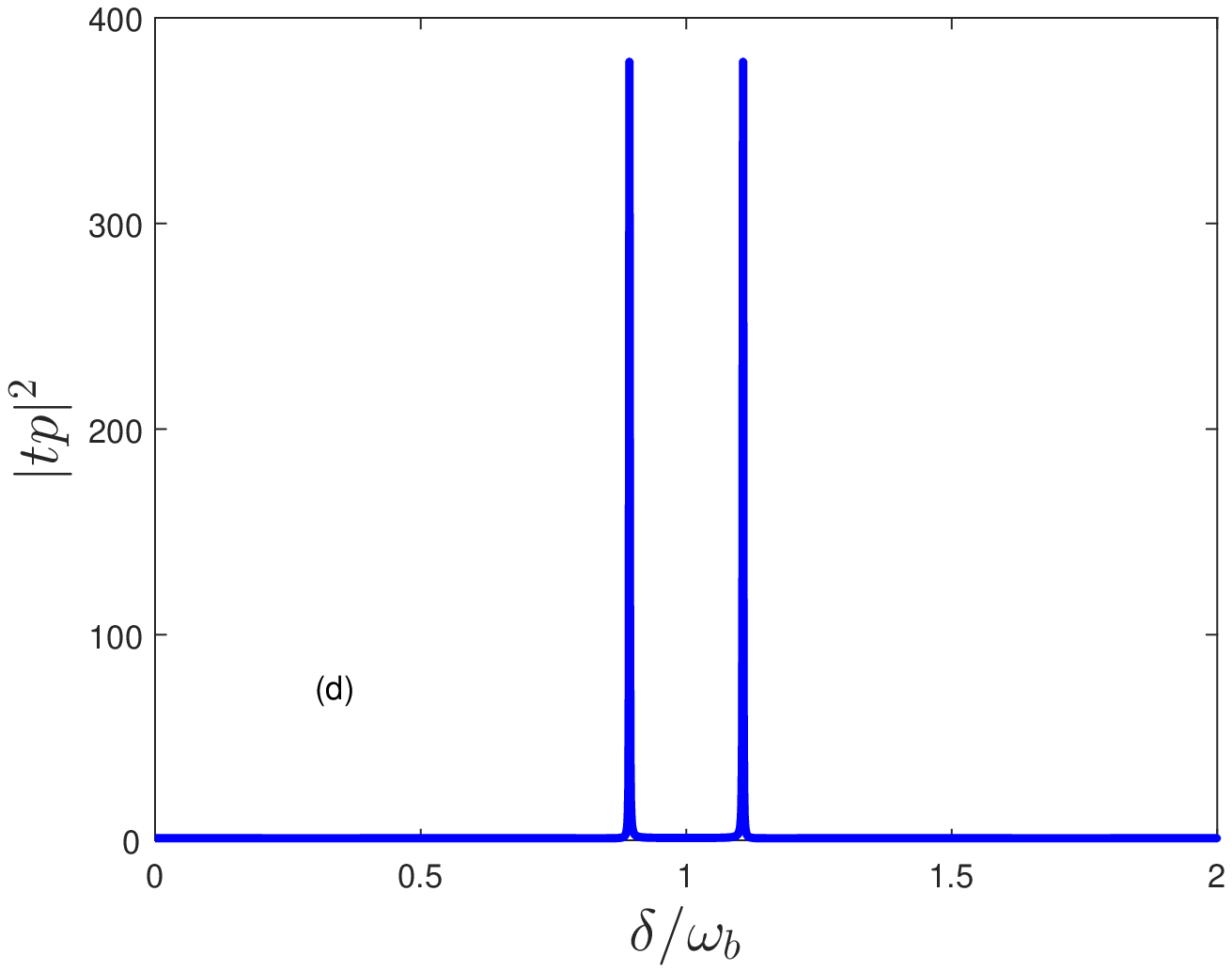}
	\hspace{0in}%
	\caption{The transmission $\left| t_p\right| ^2$  spectrum of probe field as function of $\delta/\omega_b$ when $G/2\pi=2.0\rm MHz$, $g_1/2\pi=6.0\rm MHz$ (a)  $J/2\pi=0.64\rm MHz$, (b)  $J/2\pi=0.8\rm MHz$, (c)$J/2\pi=2\rm MHz$ and (d)$J/2\pi=4\rm MHz$. }
	\label{fig4}
\end{figure}

We show the transmission spectrum when three coupling constants are nonzero simultaneously and coupling between magnon-photon lager than magnon-phonon $g_1>G$ in Fig~\ref{fig4}(a)-(d). We got only one amplification peak when the coupling between two cavities $J/2\pi=0.64\rm MHz$, another upward peak appeared with the increasing of $J$, and the height of two peaks is the same and the amplification effect induced by the interaction of magnon-phonon and magnon-photon were superior at this time, this is because magnon and phonon can also scatter the photons of active cavity field into the probe field. And since the excited states of the cavity field are pumped into higher energy levels, they stay long enough can also be amplified by stimulated radiation. Amplification area becomes wider when $J$ getting lager simultaneously. These results provide an effective way to realize continuous optical amplification and have practical significance for the construction of quantum information processing enhancement signal based on cavity magnetic system.  

\begin{figure}
	\centering
	\includegraphics[width=0.8\linewidth,height=0.2\textheight]{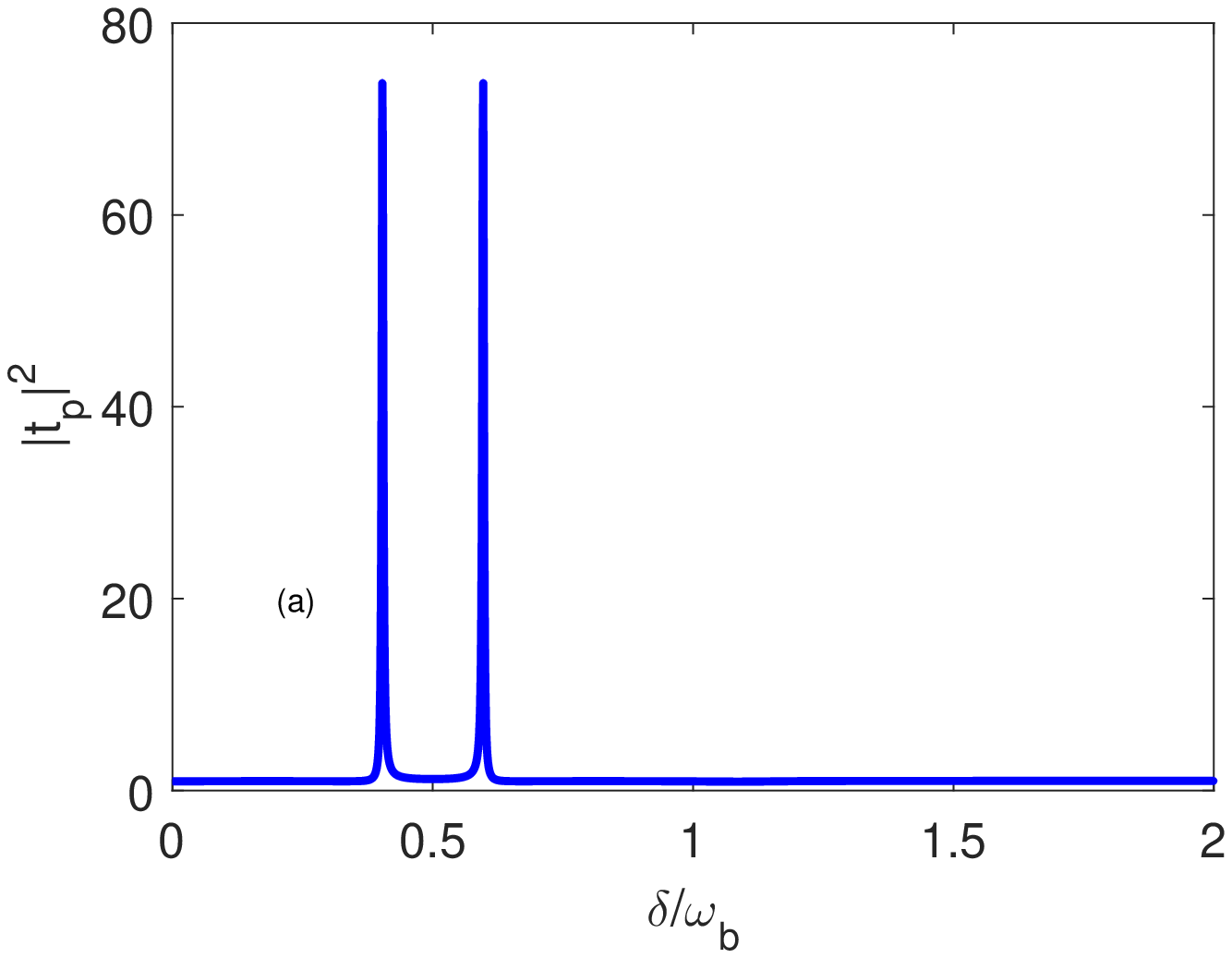}
	\hspace{0in}%
	\includegraphics[width=0.8\linewidth,height=0.2\textheight]{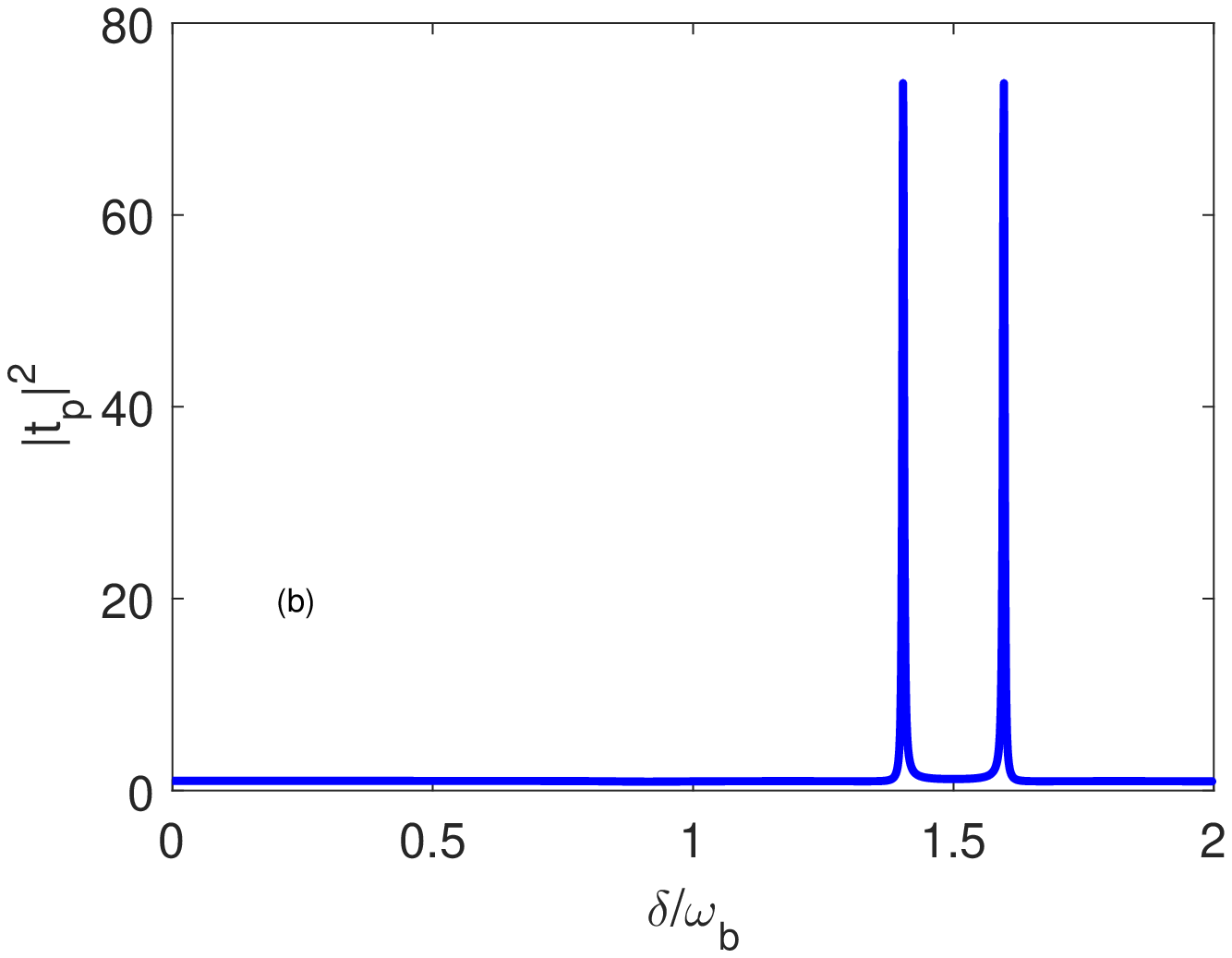}
	\hspace{0in}%
	\caption{The transmission $\left| t_p\right| ^2$  spectrum of probe field as function of $\delta/\omega_b$ when three coupling constants are nonzero, (a)  $\widetilde{\Delta}_m=0.5\omega_b$, (b)  $\widetilde{\Delta}_m=1.5\omega_b$.}
	\label{fig5}
\end{figure}

Finally, we plotted the transmission spectrum of the probe field against the scaled detuning $\delta/\omega_b$, for different values of $\widetilde{\Delta}_m$. From Fig~\ref{fig5}(a)-(d), The obvious displacement of the two peaks means that we can not only change the value of amplification and the size of the amplification region by adjusting the coupling strength, but also flexibly change the location of the amplification region.

Moreover, the phase $\phi_t$ of the output field can be given as 
\begin{equation}\label{e010}
\begin{aligned}
\phi_t&=\arg[\varepsilon_{out}]\\
\end{aligned}
\end{equation}

\begin{figure}
	\centering
	\includegraphics[width=0.8\linewidth,height=0.20\textheight]{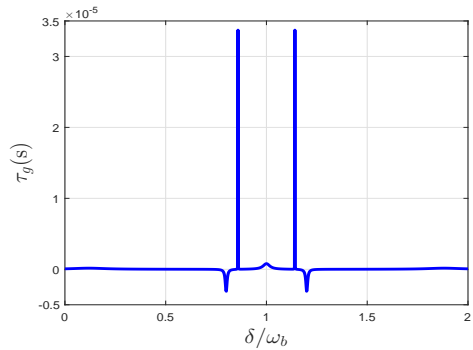}
	\hspace{0in}%
	\caption{The group delay $\tau_g$ as functions of $\delta/\omega_b$ when $G=2\rm MHz$, $J/2\pi=6.3\rm MHz$, $g_1/2\pi=6.1\rm MHz$.}
	\label{fig6}
\end{figure}

And the rapid phase dispersion of output field can cause the group delay, which can expressed as  

\begin{equation}\label{e011}
\begin{aligned}
\tau_g=\frac{\partial\phi_t}{\partial\omega_{pr}}\\
\end{aligned}
\end{equation}

 From Fig.~\ref{fig6} shows that the group delay $\tau_g$ as a function of the detuning $\delta/\omega_b$ when three coupling constants are present. We can observe double peaks and double dips, peaks corresponding positive group delay  i.e., slow light propagation, dips corresponding negative group delay means fast light propagation. And we can realize group speed delay of $3.5\times10^{-5}s$, a tunable switch from slow to fast can be achieved by adjusting the gain of active cavity or coupling constants.

\section{Conclusion}\label{s4}
In conclusion, we study the transmission of probe field in the situation of $\mathcal{PT}$-symmetric-like under a strong control field in a hybrid magnomechanical system in the microwave regime and realized ideal induced amplification when three coupling constants are nonzero simultaneously, which due to gain cavity, magnon and phonon can also scatter the photons into the dissipative cavity. Therefore, our results are not only providing rich scientific insight in terms of new physics but also potentially have important long-term technological implications, including the development of on-chip optical systems that support states of light that are immune to back scatter, are robust against perturbation and feature guaranteed unidirectional transmission. Then we achieved a group delay of $3.5\times{10}^{-5}$ seconds. Slowing down the energy speed of light allows photons to interact with matter enough to enhance some nonlinear effects. And because the dispersion curve is relatively flat, a small change in frequency will also cause a large change in photon momentum, so it can be made into a more sensitive optical switch. Finally, the slow light effect slows down the energy speed of light, which can play a role of storing photons and quantum optical chips.  \label{key}

\begin{center}
{\bf{ACKNOWLEDGMENTS}}
\end{center}
This work is supported by the National Natural Science Foundation of China (Grant No. 62165014) and the Fujian Natural Science Foundation (Grant No. 2021J01185).

\end{document}